\documentclass[aip,amsmath,amssymb,onecolumn,notitlepage,
reprint,%
]{revtex4-2}  

\usepackage{graphicx}
\usepackage{dcolumn}
\usepackage{bm}
\usepackage[utf8]{inputenc}
\usepackage[T1]{fontenc}
\usepackage{mathptmx}
\usepackage[normalem]{ulem} 	
\usepackage[usenames,dvipsnames]{color}
\usepackage{xcolor} 
\usepackage{SIunits} 
\usepackage{hyperref} 
\usepackage{babel}
\addto{\captionsenglish}{}

\newif\iffinal
\finalfalse  

\newcommand{\del}[1]{\sloppy{\textcolor{blue}{\sout{#1}}}} 

\newcommand{\new}[1]{\textcolor{black}{#1}}

\newcommand{\erms}[0]{\ensuremath{E_\mathrm{rms}}}

\iffinal
	
	\renewcommand{\del}[1]{}

	\renewcommand{\macom}[1]{}
	
\else
\fi

\makeatletter
\let\@fnsymbol\@fnsymbol@latex
\@booleanfalse\altaffilletter@sw
\makeatother
\begin{document}

\title{Mapping the positions of Two-Level-Systems on the surface of a superconducting transmon qubit}

\author{J\"urgen Lisenfeld}\thanks{Corresponding author: juergen.lisenfeld@kit.edu}
\affiliation{Physikalisches Institut, Karlsruhe Institute of Technology, 76131 Karlsruhe, Germany}
\author{Alexander K. H\"andel}
\affiliation{Physikalisches Institut, Karlsruhe Institute of Technology, 76131 Karlsruhe, Germany}
\author{Etienne Daum}
\affiliation{Physikalisches Institut, Karlsruhe Institute of Technology, 76131 Karlsruhe, Germany}
\author{Benedikt Berlitz}
\affiliation{Physikalisches Institut, Karlsruhe Institute of Technology, 76131 Karlsruhe, Germany}
\author{Alexander Bilmes}
\affiliation{Physikalisches Institut, Karlsruhe Institute of Technology, 76131 Karlsruhe, Germany}
\affiliation{Google Research, Mountain View, CA, USA  }
\author{Alexey V. Ustinov}
\affiliation{Physikalisches Institut, Karlsruhe Institute of Technology, 76131 Karlsruhe, Germany}

\date{\today \phantom{   } 
}

\begin{abstract}
	\centering\begin{minipage}{\linewidth}
		\textbf{The coherence of superconducting quantum computers is severely limited by material defects that create parasitic two-level-systems (TLS). Progress is complicated by lacking understanding how TLS are created and in which parts of a qubit circuit they are most detrimental.
        Here, we present a method to determine the individual positions of TLS at the surface of a transmon qubit.       
        We employ a set of on-chip gate electrodes near the qubit to generate local DC electric fields that are used to tune the TLS' resonance frequencies. The TLS position is inferred from the strengths at which TLS couple to different electrodes and comparing them to electric field simulations.
        We found that the majority of detectable surface-TLS was residing on the leads of the qubit's Josephson junction, despite the dominant contribution of its coplanar capacitor to electric field energy and surface area. This indicates that the TLS density is significantly enhanced near shadow-evaporated electrodes fabricated by lift-off techniques.
        Our method is useful to identify critical circuit regions where TLS contribute most to decoherence, and can guide improvements in qubit design and fabrication methods.
	}
	\end{minipage}
\end{abstract}

\maketitle 
\setlength{\parskip}{-0.25cm}

\section*{Introduction}
\begin{figure*}[ht]
\centering
    \includegraphics[width=\textwidth]{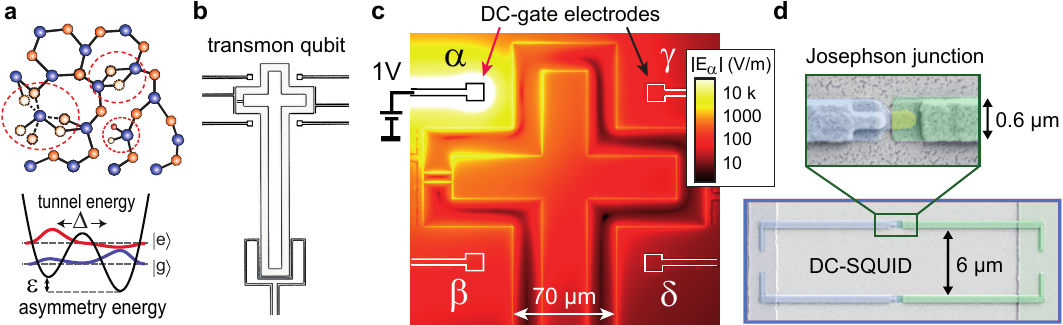}      
	\caption{\textbf{Qubit design to measure TLS locations.}
   \textbf{a} Models of Two-Level-Systems (TLS) formed by delocalized atoms in an amorphous material, and corresponding TLS double-well potential that is characterized by the tunneling energy $\Delta$ between the TLS states and the asymmetry energy $\varepsilon$.   
   \textbf{b} Layout of the transmon qubit, formed by a cross-shaped island that is connected via two Josephson junctions to the surrounding ground plane.
   \textbf{c} Four electrodes indicated by $\alpha$ to $\delta$ are placed around the qubit to generate locally concentrated DC-electric fields. The color encodes the simulated magnitude of the DC-electric field at the sample surface when 1V is applied to electrode $\alpha$.
   \textbf{d} False-colored SEM picture of the DC-SQUID and a single Josephson junction.  
   }
	\label{fig:1}
\end{figure*}

The nature of two-level tunneling systems (TLS) in amorphous materials has been puzzling generations of physicists~\cite{yu2022two}. Today, TLS are recognized as the primary source of decoherence in superconducting qubits.
A type of TLS that was well-studied in glasses is thought to originate in the tunneling of a single or a few atoms between two slightly different locations in the disordered material as illustrated in Fig.~\ref{fig:1}\textbf{a}~\cite{enss2005low}. In superconducting circuits, amorphous surface oxides on electrodes and those used for tunnel barriers of qubit junctions are thus a known host for TLS~\cite{Martinis2005, Barends13, Wang:APL:2015,Lisenfeld19, Bilmes_2020,de2021materials}. In addition, microfabrication techniques were shown to spoil the crystallinity of the substrate and to leave residuals of glassy photoresist~\cite{Quintana_2014,Dunsworth2017,gingras2025improving}. There is a variety of other models of TLS formation, and it remains unknown which types of TLS are limiting qubit coherence~\cite{muller2019towards}. To obtain insight into the elusive microscopic TLS structure,  atomistic modeling has gained in importance and was e.g. used to characterize TLS formed by Hydrogen interstitials\cite{holder2013bulk}, by dangling surface atoms~\cite{wang2025superconducting}, and by tunneling atoms in Josephson junctions~\cite{dubois20153d,paz2014}.\\

When the tunneling entity carries a charge, TLS defects possess an electric dipole moment by which they couple to the AC electric field of the resonator or qubit mode, and they quickly dissipate resonantly absorbed energy via their strong phonon coupling. Optimizing the circuit design in order to minimize the coupling of TLS to the AC-electric field from the qubit mode is thus a prerequisite for long coherence times~\cite{Wang:APL:2015,ganjam2024surpassing}.
The experimental progress currently relies on laborious experiments searching for better materials and improving fabrication procedures\cite{wang2025superconducting,murray2021material,ganjam2024surpassing,biznarova2024mitigation}.
A standard method is to extract TLS loss in resonators from their power-dependent quality factor. In qubits however, energy relaxation can be dominated by only a few of the most strongly coupled near-resonant TLS. Moreover, the resonance frequencies of TLS often fluctuate due to their electric dipole or longer-range phonon interaction with thermally activated TLS~\cite{faoro2014generalized, muller2015interacting}, due to diffusing charge~\cite{bilmes2017electronic}, and due to the impact of high-energy particles which may redistribute the states of neighboring charge traps and bi-stable TLS~\cite{mcewen2022cosmic,thorbeck2023,bertoldo2023cosmic}. The resulting fluctuations of qubit resonance frequencies and coherence times~\cite{klimov2018fluctuations, burnett2019decoherence,schloer2021} are especially problematic for quantum processors as they rely on well-calibrated and stable qubits.\\

The strong interaction of TLS with qubits allows one to characterize them individually~\cite{martinis2005decoherence,neeley2008process,Barends13}.  TLS swap spectroscopy reveals the resonance frequencies of sufficiently strongly coupled TLS by detecting minima in the qubit energy relaxation time $T_1$ that is measured as a function of qubit frequency~\cite{Barends13,osman2023mitigation,wallraff2025}. This method becomes especially powerful when it is combined with means to manipulate the TLS' properties in-situ. Tuning TLS by applied mechanical strain has revealed their interactions with coherent~\cite{Lisenfeld2015} and with thermal~\cite{meissner2018probing} TLS, and was used to characterize the TLS' coherent dynamics~\cite{lisenfeld2016decoherence}.\\
Similarly, TLS were tuned by an applied DC-electric field and individually characterized with superconducting resonators~\cite{sarabi2016projected,Hung_2022,deGraaf2021etuning}.
In qubits, E-field tuning allows one to identify whether a TLS is residing in the tunnel barrier of a qubit junction~\cite{Lisenfeld19}, and provides means to enhance qubit $T_1$ times~\cite{lisenfeld2023enhancing, chen2025etuning} and their temporal stability\cite{dane2025performance}.
When the tuning electric field can be spatially varied, e.g. by using two independently biased gate electrodes placed above and below the qubit chip, it is possible to obtain information on the circuit interface at which TLS reside\cite{Bilmes_2020}. Recently, Hegedüs et al. demonstrated scanning gate microscopy to determine the  positions and electric dipole moment orientations of individual TLS at the surface of a superconducting resonator~\cite{hegedus2024}.\\

Here, we demonstrate a method to generate maps of the locations of individual TLS on the surface of a transmon qubit. The TLS locations are inferred from their measured coupling strengths to each of four on-chip gate electrodes that are placed around the qubit, realizing a method of trilateration. The majority of all detected surface-TLS were found to reside near the leads of the Josephson junctions. Considering the dominant contribution of the qubit's planar capacitor and ground plane to surface area and electric field energy, this result indicates that the TLS density is enhanced at shadow-evaporated electrodes that are deposited by additive lift-off techniques, in contrast to subtractive etching.
\\

\section*{Methods}
\subsection*{Qubit sample}
 The transmon qubit sample, shown in Figs~\ref{fig:1}~b-d, is based on the \textit{XMon}-design by Barends et al.~\cite{Barends13} and consists of a DC-SQUID and a cross-shaped island that forms a shunt capacitor with the surrounding ground plane. In addition, the design integrates four gate electrodes labeled $\alpha...\delta$ in vicinity of the qubit island which are used to tune the TLS by local DC-electric fields. \\
 The simulated E-field strength $E_\alpha$ when a voltage of 1V is applied to the $\alpha$-electrode is shown by color in Fig.~\ref{fig:1}\,c. Due to the large spatial E-field gradient, the response of a TLS depends sensitively on its distance to the gate electrode. The TLS position can thus be estimated by measuring its tuning strengths to different electrodes and comparing them to the simulated strengths of the local electric fields.\\
 
The qubit was fabricated from aluminum on a sapphire substrate, using optical lithography and dry etching for the ground plane and qubit island, and eBeam-patterned Dolan bridges to form the tunnel junctions and their leads (shown in Fig.~\ref{fig:1}d) in a 3-angle shadow evaporation process that avoids unwanted stray junctions\cite{Bilmes_2021}. Details on sample fabrication and circuit parameters are found in~\ref{app:sample}. The sample was cooled to a temperature of 25 - 30 mK and measured in a standard setup as detailed in \ref{app:setup}.\\
The qubit showed energy relaxation times  $T_1$ between 5 to 8~\textmu s at operation frequencies between 5 and 5.5~GHz. Similarly fabricated qubits without on-chip gate electrodes achieved only slightly longer $T_1$-times between 10 and 20\,\textmu s~\cite{Bilmes_2021,lisenfeld2023enhancing}, \new{and these were mostly limited by TLS near circuit electrodes~\cite{Lisenfeld19,daum2025mergemon}}.
However, the observed $T_1$-time falls within the estimate range of the radiative loss via the capacitive coupling to the four electrodes, which were placed in close vicinity to the qubit island to enhance the spatial resolution in TLS localization. This loss channel can be mitigated with an improved design of the qubit and on-chip electrodes~\cite{bilmes2021sensors}. Further details on loss are discussed in \ref{app:loss}.\\
 
\subsection*{TLS spectroscopy}
\begin{figure*}
	\centering
	\includegraphics[width=\textwidth]{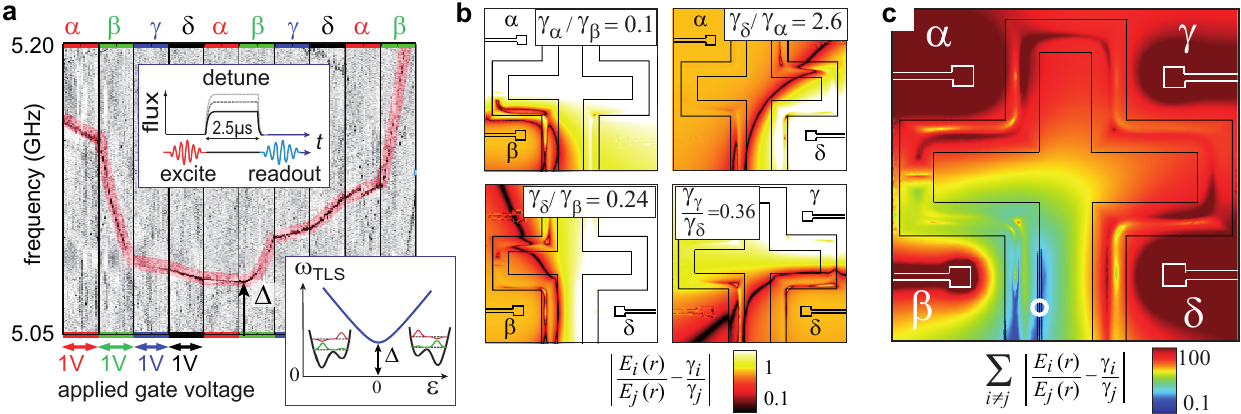}
	\caption{\textbf{Finding the location of a TLS.}
		\textbf{a}  TLS spectroscopy using the protocol in the inset, to reveal the resonance frequencies of TLS from minima  (dark pixels) in the resulting qubit population $P(|1\rangle)$.
		In each segment, the voltage on a different electrode $\alpha..\delta$ is increased by 1V. The highlighted trace shows a TLS as it is tuned through its symmetry point  (arrow) according to Eq.~(\ref{eqn:myTLShyp}) as illustrated in the lower inset. The TLS' response strengths $\gamma_i$ to the different electrodes are obtained by fitting such traces to Eq.~(\ref{eqn:myTLShyp}), and provide information about the TLS' distance to the electrodes.
		\textbf{b} 
		Difference between the measured TLS response strength ratio $\gamma_i/\gamma_j$ and the corresponding simulated E-field ratio (colorscale) for the TLS observed in \textbf{a}. Minima (dark pixels) indicate possible TLS positions. Each panel shows data from a different electrode pair as marked in the legends together with the measured tuning ratio.
		\textbf{c}
		The colorscale shows the difference sum $\sigma$ (Eq.~\ref{eqn:mydiff}) over all 6 unique combinations of electrode pairs. The minimum (white circle) marks the most probable TLS position.
	}
	\label{fig:2}
\end{figure*}

The resonance frequency of a charged TLS is given by the hyperbolic function
\begin{equation}\label{eqn:myTLShyp}
 \omega_{\mathrm{TLS}} = \sqrt{\Delta^2 + (\epsilon + 2 \mathbf{p} \cdot \mathbf{E})^2} / \hbar,
\end{equation}
where $\Delta$ is the tunneling energy between its two states, and $\epsilon$ is a background asymmetry energy of the TLS' double-well potential (see Fig.~\ref{fig:1}\,a) that depends on local static electric and strain fields. In addition, the asymmetry energy is tuned by the component of the applied E-field $\textbf{E}$ at the position of the TLS that is parallel to the TLS' electric dipole moment~$\mathbf{p}$. In our experiments, the E-field $\mathbf{E = E_\alpha + E_\beta + E_\gamma + E_\delta}$ is controlled via the applied voltages $V_\alpha...V_\delta$ on the four gate electrodes.\\
We detect TLS resonances using the TLS swap-spectroscopy protocol~\cite{Barends13,Lisenfeld2015,Lisenfeld19} depicted in the top inset of Fig.~\ref{fig:2}a. The qubit is excited by a microwave $\pi$-pulse and tuned to one of various probe frequencies for a duration of $2.5$\,\textmu s to allow for interactions with TLS. The remaining qubit excitation $P_{|1\rangle}$ then provides an estimate for the qubit $T_1$ time at the probe frequency\cite{lisenfeld2023enhancing}, which shows a minimum when the qubit is in resonance with a sufficiently strongly coupled TLS.\\
 
To measure the TLS' coupling strengths to the four DC-electrodes, their resonances are traced by TLS spectroscopy while the voltages on the DC-electrodes are swept. As an example, Figure~\ref{fig:2}a shows the resonance of a TLS that was tuned through the symmetry point of its hyperbola Eq.~(\ref{eqn:myTLShyp}), where in each segment the voltage on the indicated gate electrode $\alpha...\delta$ was stepwise increased by 1\,V while other voltages were kept constant.
The slope of the hyperbola in the different segments then depends on the tuning strength of the TLS by the corresponding electrode. It is obtained by fitting such traces to Eq.~(\ref{eqn:myTLShyp}), where the factor for the induced asymmetry energy $2\textbf{p}\cdot \textbf{E}$ is replaced by $\sum_i \gamma_i V_i$. Here, $i  \in \{\alpha, \beta, \gamma, \delta\}$ indicates the electrode that is biased by the voltage $V_i$, and $\gamma_i$ are the fitting tuning strengths that contain information on the distance of the TLS to the corresponding electrodes.\\

\subsection*{TLS localization}

Possible locations of the TLS could in principle be inferred from the measured tuning strengths by searching for positions (x,y) where the simulated electric field $\mathbf{E}_i (x,y)$ fulfills the equation $2\mathbf{p}\cdot \mathbf{E}_i(x,y,V_i) = \gamma_i V_i$.
However, this would require knowledge of the TLS' electric dipole moment $\mathbf{p}$ and its orientation relative to the local E-field.
In our analysis, we therefore consider only relative tuning strengths of the TLS by different electrodes, and search for positions (x,y) fullfilling the equations
\begin{equation}
    \frac{2\mathbf{p}\cdot\mathbf{E_i}(x,y)}{2\mathbf{p}\cdot\mathbf{E_j}(x,y)} = \frac{\gamma_i V_i}{\gamma_j V_j} \mathrm{,\ \ where\ \ } \{i \neq j\} \in {\alpha, \beta, \gamma, \delta}.
    \label{eqn:myratio}
\end{equation}
In these equations, the TLS' electric dipole moment $\mathbf{p}$ can be eliminated given that the two fields $\mathbf{E_i}$ and $\mathbf{E_j}$ have parallel orientation at the position of the TLS. In the following, we argue that this is indeed the case in our experiment, because we are detecting only TLS in close vicinity to the edges of qubit electrodes where all electric fields are sufficiently aligned.\\

\begin{figure*}
    \centering
    \includegraphics[width=\textwidth]{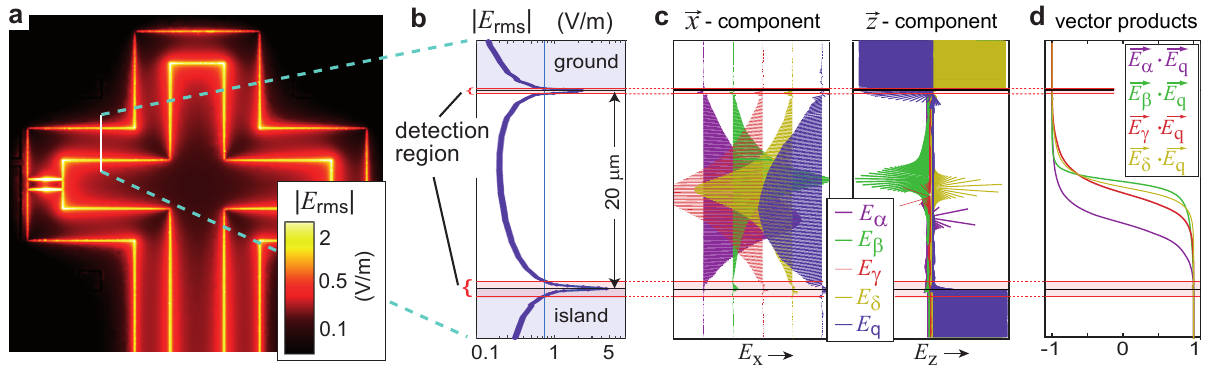}
    \caption{\textbf{Strength and orientations of AC- and DC-electric fields.}    	
    	\textbf{a} Magnitude of the qubit's AC-electric field $|\erms|$ as simulated with Ansys HFSS. \textbf{b} Cross-section of $|\erms|$ along the white line in \textbf{a}. TLS can only be detected in the red shaded area where the field exceeds a minimum strength $E_\mathrm{min}$ (blue vertical line). \textbf{c} Components of the electric fields from the four electrodes and the qubit, plotted in $\vec{x}-$ and $\vec{z}$ directions (left and right panel). Near the center of the gap between ground plane and qubit island, the fields induced by the gate electrodes change their direction and point in different directions. \textbf{d} The normalized vector product between the fields of gate electrodes and the qubit approaches unity near the electrode edges where all fields point in the same direction.}
    \label{fig:3}
\end{figure*}

To be able to detect a TLS in qubit $T_1$-swap-spectroscopy, its (resonant) coupling strength to the qubit $g=\left(\frac{\Delta}{\hbar \omega_{\mathrm{TLS}} }\right)\  \mathbf{p}\cdot\mathbf{\erms}$ must be large enough to result in a measurable decrease of the qubit's energy relaxation rate.
Thus, TLS can only be detected in circuit regions where the qubit's AC-electric field strength $|\erms|$ exceeds $g_\mathrm{min}/ p_\parallel$.
Assuming $\Delta/\hbar\omega_\mathrm{TLS} \approx 1$ (most strongly coupling TLS near their symmetry point) and a field-parallel dipole moment of $p_\parallel \approx 1 e$\AA~\cite{Martinis2005,Barends13,carruzzo2021,Hung_2022}, the required minimum coupling strength $g_\mathrm{min}$ can be estimated from the energy relaxation rate of the resonantly coupled qubit-TLS system~\cite{Barends13} $\Gamma_1 = 2\,(g_\mathrm{min}/\hbar)^2 / \Gamma + \Gamma_{1,Q}$, where $\Gamma_{1,Q}$ is the energy relaxation rate of the isolated qubit, and $\Gamma = \Gamma_{1,\mathrm{TLS}}/2 + \Gamma_{2,\mathrm{TLS}} + \Gamma_{1,\mathrm{Q}}/2 + \Gamma_{2,\mathrm{Q}}$ is the sum of TLS and qubit energy relaxation and dephasing rates. 
The assumption that TLS are detected if they reduce the qubit's $T_1$ time by a factor of $\kappa$ translates into a minimum coupling strength of $g_\mathrm{min} = \hbar \sqrt{\kappa \cdot \Gamma_\mathrm{1,Q} \,\Gamma /2}$ and corresponding minimum AC-electric qubit field strength $E_\mathrm{min} = g_\mathrm{min} / p_\parallel$.
A plot of these relations can be found in \ref{app:minfield}. For a qubit $T_1=7\ $\textmu s, TLS coherence times in the range of $T_\mathrm{1,TLS} \approx T_\mathrm{2,TLS}  \approx 0.1 - 2\,$\textmu s \cite{Barends13,Shalibo2010,Lisenfeld19,Lisenfeld2015,chen2024,weeden2025statistics}, and $\kappa = 5\%$, we find a $E_\mathrm{min}$ range of $\approx 0.3 - 3$ V/m. \\ 

Figure~\ref{fig:3}a shows the simulated magnitude of the qubit's AC-electric field $|\erms|$. A cross section through the edges of ground plane and qubit island is shown in Fig~\ref{fig:3}b and illustrates that in our qubit sample, TLS can only be detected within an $\approx 1 - 2$ \textmu m distance from the edge of qubit electrodes where $|\erms| > E_\mathrm{min}$. Since $E_\mathrm{min}$ decreases with the square root of the qubit's $T_1$ time, more coherent qubits are affected by TLS in a wider area which includes more weakly coupled ones as illustrated in~\ref{app:minfield}.\\

To show that the electric fields from different on-chip electrodes are indeed sufficiently parallel, such that the dipole moment in Eq.~(\ref{eqn:myratio}) can be canceled in order to simplify the solution for possible TLS positions, in Fig.~\ref{fig:3}c we plot the electric fields' $\vec{x}-$ and $\vec{z}$-components along the white line in Fig.~\ref{fig:3}a. Near the center of the gap between ground and qubit island, the E-fields point in different directions. However, within the short distance to the electrode edges where TLS are detected in our experiment, all E-fields are well aligned since their vector products with the qubit field $\vec{E_q}$ approach unity as plotted in Fig.~\ref{fig:3}d. \new{This is also the case near the junction leads as shown in~\ref{app:minfield}.}\\

After dropping the TLS' electric dipole moment $\mathbf{p}$ in the vector product in Eq.~(\ref{eqn:myratio}), we find the most probable TLS position $(x,y)$ by minimizing the sum of residuals
\begin{equation}
    \sigma  = \sum_{i \neq j} \left |\frac{E_i(x,y)}{E_j(x,y)} - \frac{\gamma_i}{\gamma_j} \right |\mathrm{,\ \ where\ \ } \{i \neq j\} \in {\alpha, \beta, \gamma, \delta}.
    \label{eqn:mydiff}
\end{equation}

Each of the six summands in Eq.~(\ref{eqn:mydiff}) is the difference between the measured tuning ratio of two electrodes with the corresponding simulated E-field ratio at position $(x,y)$. Figure~\ref{fig:2}b shows exemplarily the contribution of four summands for the TLS observed in Fig.~\ref{fig:2}a, where minima (dark pixels) indicate possible TLS positions that are confined along approximate circles centered at the electrode which is nearest to the TLS. For example, the TLS shown in Fig.~\ref{fig:2}a shows a 10 times stronger response to the $\beta$-electrode than the $\alpha$-electrode. This places possible TLS positions in vicinity of the $\beta$-electrode as shown in the top left panel of Fig.~\ref{fig:2}b.\\

Figure~\ref{fig:2}c shows a plot of the complete difference sum of Eq.~(\ref{eqn:mydiff}) that contains information from all six electrode pairs, and where the global minimum (marked by a white circle) then indicates the most likely TLS position. In this analysis, we allow only solutions of TLS positions within the region where the qubit's AC field is strong enough ($|\erms| > E_\mathrm{min}$) so that TLS can be detected.\\
\begin{figure*}
    \centering
    \includegraphics[width=\textwidth]{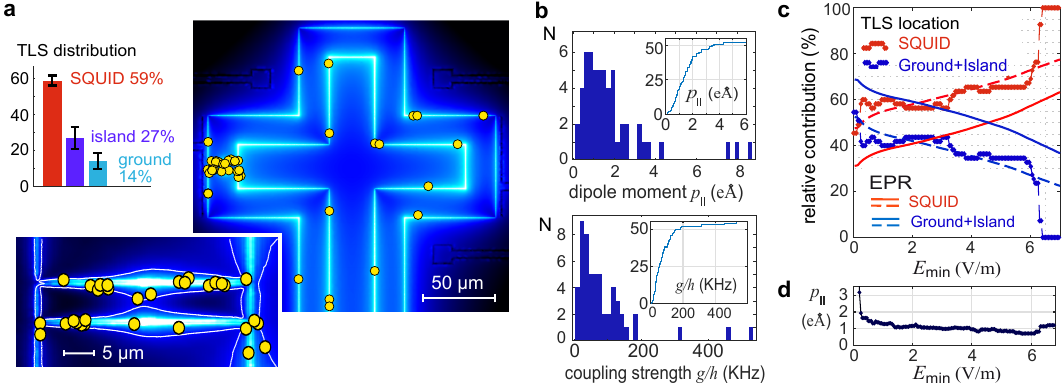}
    \caption{\textbf{TLS locations and properties.}
    	\textbf{a} Map of the individual positions of detected surface-TLS (yellow circles). Most TLS ($\approx$~59\%) were found at the Josephson junctions leads (see lower inset for a zoom on the DC-SQUID). Near the edges of capacitor island and ground plane, 27\% and 14\% of TLS were identified, respectively. 
    \textbf{b} Histograms and cumulative distributions (insets) of the TLS' electric dipole moment component $p_\parallel$ (upper panel) and TLS-qubit coupling strengths $g/h$ (lower panel) as estimated from the TLS' positions, their coupling strengths to the electrodes, and the local magnitude of the qubit AC field \erms.
    \textbf{c} Percentage of TLS identified on the SQUID vs. ground plane plus island, plotted as a function of the electric field threshold $E_\mathrm{min}$ that accounts for TLS observability in swap spectroscopy. The solid lines are the \new{energy participation ratios (EPRs) of the SQUID (red) and the ground plus island (blue) in regions where} $E_\mathrm{rms}>E_\mathrm{min}$. The dashed lines are a fit of these ratios to the observed TLS distribution, obtained by assuming that the TLS density in the SQUID area is enhanced by a factor of 2.         
    \textbf{d} Determined median TLS electric dipole moment $p_\parallel$ as a function of $E_\mathrm{min}$.   
	}
    \label{fig:4}
\end{figure*}
\section*{Results}
\subsection*{TLS distribution}
To generate a map of the individual TLS positions on the qubit circuit, we swept the voltages on each of the four electrodes in a range between -60V and +100V and performed TLS spectroscopy in a 150-MHz wide window as shown in Fig.~\ref{fig:2}a.
In total, 55 TLS were observed in a single qubit sample whose tuning strengths to all four electrodes could be characterized (see \ref{app:data} for further details).
\new{For the analysis, TLS were discarded whose coupling strength to the upper electrodes $\alpha$ and $\gamma$ were indistinguishable from zero so that their location could not be precisely determined. This is the case for TLS that reside on the long arm of the qubit island (see Fig.~\ref{fig:1}b).}\\ 
\indent The determined individual TLS positions are marked by yellow circles in Fig~\ref{fig:4}a. The majority of TLS (58\%) were found to reside at the leads of the qubit's Josephson junctions. Near the edges of the qubit island and the ground plane, 25\% and 16\% were found, respectively.
In this example, a relatively small value of $E_\mathrm{min}=0.75$~V/m was chosen as the threshold field above which TLS can be detected, which limits the area of allowed TLS positions to the thin white line in the inset of Fig.~\ref{fig:4}a.\\
\indent To check the mapping procedure's reliability, the analysis is repeated for various values of $E_\mathrm{min}$. The resulting percentages of TLS found on the SQUID leads vs. those on the qubit island and ground plane are plotted in Fig.~\ref{fig:4}c. These remain at a ratio of 60:40 within the most probable range of $E_\mathrm{min} \approx 0.3..3$~V/m as estimated above, which backs the robustness of the method. Above a value of $E_\mathrm{min}\approx 6$~V/m, the procedure fails since it is forced to place all TLS on the DC-SQUID. \new{For the given data and qubit sample, we estimate an uncertainty in the TLS positions of about 6~\textmu m as further discussed in~\ref{app:resolution}. This resolution can be enhanced with more suitable qubit designs that avoid sharp corners.}\\

\subsection*{Excess TLS density near the SQUID}
Notably, the observed TLS distribution points towards an excess density of TLS near the DC-SQUID leads.
Since the probability to detect a surface-TLS scales with the square of the local qubit field, the expected ratio of the numbers of TLS observed near the SQUID vs. ground plane and qubit island can be estimated by comparing \new{their energy participation ratios (EPRs)}.
By calculating the EPR, we limit the integrals over $|\erms|^2$ to regions where $E_\mathrm{rms}>E_\mathrm{min}$ to account for TLS detectability in swap spectroscopy, which sets this calculation apart from the common participation ratio analysis~\cite{Wang:APL:2015} that includes also weakly coupled TLS. \new{Further details are given in \ref{app:minfield}.} The result is plotted with solid red and blue lines in Fig.~\ref{fig:4}c, and predicts that most TLS (>50\%) would be observed on either the qubit island or ground plane in the whole range of reasonable values $E_\mathrm{min}< 4$\,V/m, in stark contrast to the experimental data.\\
\indent This finding can be reconciled assuming that the TLS density near the DC SQUID is about two times larger than near qubit island and ground plane, presumably due to its different fabrication procedure. 
The expected TLS distribution for this case is shown by the dashed red and blue lines in Fig.~\ref{fig:4}c, which are obtained by scaling the electric field energy $E_\mathrm{rms}^2$ integrated over the SQUID area by a factor of two. We find a very good agreement with the experimental result that is mostly insensitive to the choice of $E_\mathrm{min}$ in the mapping algorithm. \\

From the TLS' estimated positions, we can calculate their electric dipole moments using the measured tuning strengths $\gamma_i$ and the simulated local E-fields of corresponding DC-gate electrodes. Similarly, the TLS' coupling strength to the qubit $g$ can be estimated using the simulation of the qubit's AC-electric field strength at the TLS' position. Figure~\ref{fig:4}b shows their histograms and cumulative distributions for the representative example shown in Fig.~\ref{fig:4}a. The extracted median TLS dipole moment varies weakly with $E_\mathrm{min}$ as shown in Fig.~\ref{fig:4}d, and is estimated to $p_\parallel \approx 1.12 \pm 0.12\, e$\AA\, which is well in accordance with results obtained using other methods in qubits and lumped-element resonators~\cite{Martinis2005,Barends13,Hung_2022}.
 
\section*{Discussion}
We have demonstrated a method to determine the individual positions and electric dipole moments of TLS defects in a transmon qubit. 
The majority ($\approx 58\%$) of observed strongly interfering TLS in the qubit sample were found to reside on the qubit's DC-SQUID.
This confirms that the leads of tunnel junctions, due to their large electric energy participation, are a critical component which can dominate qubit loss~\cite{deng2023titanium,smirnov2024wiring,weeden2025statistics}. This advocates  for the use of wire-tapering techniques to dilute the electric field~\cite{martinis2021}, \new{and to minimize the size of lift-off structures.}\\

Our results additionally indicate that the TLS density near the junction leads is enhanced by a factor of $\approx 2$. This may be attributed to their different fabrication technique as compared to the qubit capacitor, which can promote TLS formation in various ways. 
For example, junction formation by shadow evaporation and electron-beam lithography is associated with larger amounts of resist residuals and enhanced roughness of junction lead interfaces~\cite{Dunsworth2017,Moskalev2022,gingras2025improving}.
The junctions were fabricated with a lift-off process which reportedly leaves excess residues~\cite{Quintana_2014,smirnov2024wiring}. Also, thinner films showed a larger density of grain boundaries associated with enhanced oxygen diffusion~\cite{biznarova2024mitigation}.\\

The technique to find TLS positions works with various (charge-resilient) qubit types such as flux and phase or transmon qubits, and provides information on the local TLS density in a single sample without the need to average over a large ensemble of differently designed qubits.
It can be applied to arbitrary qubit designs when the grid of DC-electrodes is patterned on a wafer placed above the qubits in a flip-chip configuration. This is similar to the recently demonstrated scanning gate spectroscopy~\cite{hegedus2024}, but does not require mechanical control.\\
\indent Control over the local DC-electric field also enables one to actively suppress decoherence by tuning dominating TLS defects out of the qubit resonance~\cite{lisenfeld2023enhancing,chen2025etuning,dane2025performance}. For this, multiple electrodes provide independent control over TLS at different locations which enhances the ability to decouple the qubit from the decohering bath.\\

\new{\indent The current method cannot determine at which interfaces TLS are residing. This is possible with additional gate electrode placed above and below the qubit chip, allowing one to also distinguish TLS located at the metal-substrate interface from those at the metal-air interface and the substrate as we showed in previous work~\cite{Bilmes_2020}. Moreover, the combination with TLS tuning by applied mechanical strain provides additional information of TLS densities in tunnel junction barriers~\cite{Lisenfeld19}, and is useful to enhance the numbers of observable TLS in a given qubit circuit.\\}

Our method opens door to study TLS formation due to fabrication techniques and contaminants, for example by comparing TLS densities in differently processed areas of the same qubit circuit. This approach  can serve to guide improvements in qubit fabrication and design, which is vitally needed for the advancement of large-scale superconducting quantum processors where TLS defects present a major obstacle.\\

\section*{Data availability}
The raw data that support the findings of this study are available from the public repository Zenodo, DOI:10.5281/zenodo.18847452.

\section*{Acknowledgements}
We thank Hannes Rotzinger for his support with the experimental setup and fruitful discussions, and Lukas Radtke for his commitment to the clean-room and help with sample fabrication.
This work was funded by Google, which is gratefully acknowledged. 

\section*{Author contribution}
J.L. conceived and supervised the project and wrote the manuscript. E-field tuning techniques were developed by J.L. and A.B. Data was acquired and analyzed by J.L. with support from A.K.H. and B.B. The sample was fabricated by A.K.H. and A.B. Simulations were done by A.K.H., E.D, and B.B. The experimental infrastructure was provided by A.V.U. All authors contributed to discussion and the final manuscript.

\section*{Competing interests}
The authors declare no competing financial or non-financial interests.\\

\section*{References}
\bibliographystyle{unsrt}
\bibliography{BiblioClean}

@article{yu2022two,
  title={Two-level systems and the tunneling model: A critical view},
  author={Clare, C Yu and Carruzzo, Herv{\'e} M},
  journal={Low-temperature Thermal And Vibrational Properties Of Disordered Solids: A Half-century Of Universal" Anomalies" Of Glasses},
  pages={113},
  year={2022},
  publisher={World Scientific}
}

@article{gingras2025improving,
	title={Improving Transmon Qubit Performance with Fluorine-based Surface Treatments},
	author={Gingras, Michael A and Niedzielski, Bethany M and Grossklaus, Kevin A and Miller, Duncan and Contipelli, Felipe and Azar, Kate and Burkhart, Luke D and Calusine, Gregory and Davis, Daniel and Pi{\~n}ero, Ren{\'e}e DePencier and others},
	journal={arXiv preprint arXiv:2507.08089},
	year={2025}
}

@article{carruzzo2021,
  title={Distribution of two-level system couplings to strain and electric fields in glasses at low temperatures},
  author={Carruzzo, Herve M and Bilmes, Alexander and Lisenfeld, J{\"u}rgen and Yu, Zheng and Wang, Bu and Wan, Zhongyi and Schmidt, JR and Yu, Clare C},
  journal={Physical Review B},
  volume={104},
  number={13},
  pages={134203},
  year={2021},
  publisher={APS}
}

@inbook{enss2005low,
	title={Low-temperature physics},
	author={Enss, Christian and Hunklinger, Siegfried},
	year={2005},
	publisher={Springer},
	chapter = {9},
	address = {Berlin, Heidelberg}
}

@article{wallraff2025,
	title={Mitigating losses of superconducting qubits strongly coupled to defect modes},
	author={Colao Zanuz, Dante and Ficheux, Quentin and Michaud, Laurent and Orekhov, Alexei and Hanke, Kilian and Flasby, Alexander and Bahrami Panah, Mohsen and Norris, Graham J and Kerschbaum, Michael and Remm, Ants and others},
	journal={Physical Review Applied},
	volume={23},
	number={4},
	pages={044054},
	year={2025},
	publisher={APS}
}

@article{chen2024,
  title={Phonon engineering of atomic-scale defects in superconducting quantum circuits},
  author={Chen, Mo and Owens, John Clai and Putterman, Harald and Sch{\"a}fer, Max and Painter, Oskar},
  journal={Science Advances},
  volume={10},
  number={37},
  pages={eado6240},
  year={2024},
  publisher={American Association for the Advancement of Science}
}

@article{deng2023titanium,
	title={Titanium nitride film on sapphire substrate with low dielectric loss for superconducting qubits},
	author={Deng, Hao and Song, Zhijun and Gao, Ran and Xia, Tian and Bao, Feng and Jiang, Xun and Ku, Hsiang-Sheng and Li, Zhisheng and Ma, Xizheng and Qin, Jin and others},
	journal={Physical Review Applied},
	volume={19},
	number={2},
	pages={024013},
	year={2023},
	publisher={APS}
}

@article{muller2019towards,
  title={Towards understanding two-level-systems in amorphous solids: insights from quantum circuits},
  author={M{\"u}ller, Clemens and Cole, Jared H and Lisenfeld, J{\"u}rgen},
  journal={Reports on Progress in Physics},
  volume={82},
  number={12},
  pages={124501},
  year={2019},
  publisher={IOP Publishing}
}

@article{ganjam2024surpassing,
  title={Surpassing millisecond coherence in on chip superconducting quantum memories by optimizing materials and circuit design},
  author={Ganjam, Suhas and Wang, Yanhao and Lu, Yao and Banerjee, Archan and Lei, Chan U and Krayzman, Lev and Kisslinger, Kim and Zhou, Chenyu and Li, Ruoshui and Jia, Yichen and others},
  journal={Nature Communications},
  volume={15},
  number={1},
  pages={3687},
  year={2024},
  publisher={Nature Publishing Group UK London}
}

@article{smirnov2024wiring,
	title={Wiring surface loss of a superconducting transmon qubit},
	author={Smirnov, Nikita S and Krivko, Elizaveta A and Solovyova, Anastasiya A and Ivanov, Anton I and Rodionov, Ilya A},
	journal={Scientific Reports},
	volume={14},
	number={1},
	pages={7326},
	year={2024},
	publisher={Nature Publishing Group UK London}
}

@article{bertoldo2023cosmic,
  title={Cosmic muon flux attenuation methods for superconducting qubit experiments},
  author={Bertoldo, Elia and P{\'e}rez S{\'a}nchez, Victor and Martinez, Maria and Martinez, Manel and Khalife, Hawraa and Forn Diaz, Pol},
  journal={New Journal of Physics},
  year={2023}
}

@article{mcewen2022cosmic,
  title={Resolving catastrophic error bursts from cosmic rays in large arrays of superconducting qubits},
  author={McEwen, Matt and Faoro, Lara and Arya, Kunal and Dunsworth, Andrew and Huang, Trent and Kim, Seon and Burkett, Brian and Fowler, Austin and Arute, Frank and Bardin, Joseph C and others},
  journal={Nature Physics},
  volume={18},
  number={1},
  pages={107--111},
  year={2022},
  publisher={Nature Publishing Group UK London}
}

@article{thorbeck2023,
  title={Two-level-system dynamics in a superconducting qubit due to background ionizing radiation},
  author={Thorbeck, Ted and Eddins, Andrew and Lauer, Isaac and McClure, Douglas T and Carroll, Malcolm},
  journal={PRX Quantum},
  volume={4},
  number={2},
  pages={020356},
  year={2023},
  publisher={APS}
}

@article{Wang:APL:2015,
	Author = {Wang, C. and Axline, C. and Gao, Y. Y. and Brecht, T. and Chu, Y. and Frunzio, L. and Devoret, M. H. and Schoelkopf, R. J.},
	Journal = {Appl. Phys. Lett.},
	Number = {16},
	Title = {Surface participation and dielectric loss in superconducting qubits},
	Volume = {107},
	Year = {2015}}

@article{de2021materials,
  title={Materials challenges and opportunities for quantum computing hardware},
  author={De Leon, Nathalie P and Itoh, Kohei M and Kim, Dohun and Mehta, Karan K and Northup, Tracy E and Paik, Hanhee and Palmer, BS and Samarth, Nitin and Sangtawesin, Sorawis and Steuerman, David W},
  journal={Science},
  volume={372},
  number={6539},
  pages={eabb2823},
  year={2021},
  publisher={American Association for the Advancement of Science}
}

@article{faoro2014generalized,
	title={Interacting tunneling model for two-level systems in amorphous materials and its predictions for their dephasing and noise in superconducting microresonators},
	author={Faoro, Lara and Ioffe, Lev B},
	journal={Physical Review B},
	volume={91},
	number={1},
	pages={014201},
	year={2015},
	publisher={APS}
}

@article{muller2015interacting,
  title={Interacting two-level defects as sources of fluctuating high-frequency noise in superconducting circuits},
  author={M{\"u}ller, Clemens and Lisenfeld, J{\"u}rgen and Shnirman, Alexander and Poletto, Stefano},
  journal={Physical Review B},
  volume={92},
  number={3},
  pages={035442},
  year={2015},
  publisher={APS}
}

@article{wang2025superconducting,
  title={Why superconducting Ta qubits have fewer tunneling two-level systems at the vacuum-oxide interface than Nb qubits},
  author={Wang, Zhe and Yu, Clare C and Wu, Ruqian},
  journal={Physical Review Applied},
  volume={23},
  number={2},
  pages={024017},
  year={2025},
  publisher={APS}
}

@article{klimov2018fluctuations,
  title={Fluctuations of energy-relaxation times in superconducting qubits},
  author={Klimov, Paul V and Kelly, Julian and Chen, Zijun and Neeley, Matthew and Megrant, Anthony and Burkett, Brian and Barends, Rami and Arya, Kunal and Chiaro, Ben and Chen, Yu and others},
  journal={Physical review letters},
  volume={121},
  number={9},
  pages={090502},
  year={2018},
  publisher={APS}
}

@article{Lisenfeld19,
   title={Electric field spectroscopy of material defects in transmon qubits},
   volume={5},
   ISSN={2056-6387},
   url={http://dx.doi.org/10.1038/s41534-019-0224-1},
   DOI={10.1038/s41534-019-0224-1},
   number={1},
   journal={npj Quantum Information},
   publisher={Springer Science and Business Media LLC},
   author={Lisenfeld, Jürgen and Bilmes, Alexander and Megrant, Anthony and Barends, Rami and Kelly, Julian and Klimov, Paul and Weiss, Georg and Martinis, John M. and Ustinov, Alexey V.},
   year={2019},
   month=nov }

@article{lisenfeld2016decoherence,
	title={Decoherence spectroscopy with individual two-level tunneling defects},
	author={Lisenfeld, J{\"u}rgen and Bilmes, Alexander and Matityahu, Shlomi and Zanker, Sebastian and Marthaler, Michael and Schechter, Moshe and Sch{\"o}n, Gerd and Shnirman, Alexander and Weiss, Georg and Ustinov, Alexey V},
	journal={Scientific reports},
	volume={6},
	number={1},
	pages={23786},
	year={2016},
	publisher={Nature Publishing Group UK London}
}

@article{Bilmes_2020,
   title={Resolving the positions of defects in superconducting quantum bits},
   volume={10},
   ISSN={2045-2322},
   url={http://dx.doi.org/10.1038/s41598-020-59749-y},
   DOI={10.1038/s41598-020-59749-y},
   number={1},
   journal={Scientific Reports},
   publisher={Springer Science and Business Media LLC},
   author={Bilmes, Alexander and Megrant, Anthony and Klimov, Paul and Weiss, Georg and Martinis, John M. and Ustinov, Alexey V. and Lisenfeld, Jürgen},
   year={2020},
   month=feb }

@article{weeden2025statistics,
   	title={Statistics of Strongly Coupled Defects in Superconducting Qubits},
   	author={Weeden, S and Harrison, DC and Patel, S and Snyder, M and Blackwell, EJ and Spahn, G and Abdullah, S and Takeda, Y and Plourde, BLT and Martinis, JM and others},
   	journal={arXiv preprint arXiv:2506.00193},
   	year={2025}
   }

@article{Martinis2005,
  title = {Decoherence in Josephson Qubits from Dielectric Loss},
  author = {Martinis, John M. and Cooper, K. B. and McDermott, R. and Steffen, Matthias and Ansmann, Markus and Osborn, K. D. and Cicak, K. and Oh, Seongshik and Pappas, D. P. and Simmonds, R. W. and Yu, Clare C.},
  journal = {Phys. Rev. Lett.},
  volume = {95},
  issue = {21},
  pages = {210503},
  numpages = {4},
  year = {2005},
  month = {Nov},
  publisher = {American Physical Society},
  doi = {10.1103/PhysRevLett.95.210503},
  url = {https://link.aps.org/doi/10.1103/PhysRevLett.95.210503}
}

@article{holder2013bulk,
  title={Bulk and surface tunneling hydrogen defects in alumina},
  author={Holder, Aaron M and Osborn, Kevin D and Lobb, CJ and Musgrave, Charles B},
  journal={Physical review letters},
  volume={111},
  number={6},
  pages={065901},
  year={2013},
  publisher={APS}
}

@article{murray2021material,
  title={Material matters in superconducting qubits},
  author={Murray, Conal E},
  journal={Materials Science and Engineering: R: Reports},
  volume={146},
  pages={100646},
  year={2021},
  publisher={Elsevier}
}

@article{Dunsworth2017,
  title={Characterization and reduction of capacitive loss induced by sub-micron Josephson junction fabrication in superconducting qubits},
  author={Dunsworth, A and Megrant, A and Quintana, C and Chen, Zijun and Barends, R and Burkett, B and Foxen, B and Chen, Yu and Chiaro, B and Fowler, A and others},
  journal={Applied Physics Letters},
  volume={111},
  number={2},
  year={2017},
  publisher={AIP Publishing}
}

@article{paz2014,
  title={Identification of structural motifs as tunneling two-level systems in amorphous alumina at low temperatures},
  author={Paz, Alejandro P{\'e}rez and Lebedeva, Irina V and Tokatly, Ilya V and Rubio, Angel},
  journal={Physical Review B},
  volume={90},
  number={22},
  pages={224202},
  year={2014},
  publisher={APS}
}

@article{dubois20153d,
  title={Delocalized oxygen as the origin of two-level defects in Josephson junctions},
  author={DuBois, Timothy C and Per, Manolo C and Russo, Salvy P and Cole, Jared H},
  journal={Physical review letters},
  volume={110},
  number={7},
  pages={077002},
  year={2013},
  publisher={APS}
}

@article{deGraaf2021etuning,
  title={Quantifying dynamics and interactions of individual spurious low-energy fluctuators in superconducting circuits},
  author={De Graaf, SE and Mahashabde, Sumedh and Kubatkin, SE and Tzalenchuk, A Ya and Danilov, AV},
  journal={Physical Review B},
  volume={103},
  number={17},
  pages={174103},
  year={2021},
  publisher={APS}
}

@article{schloer2021,
  title = {Correlating Decoherence in Transmon Qubits: Low Frequency Noise by Single Fluctuators},
  author = {Schl\"or, Steffen and Lisenfeld, J\"urgen and M\"uller, Clemens and Bilmes, Alexander and Schneider, Andre and Pappas, David P. and Ustinov, Alexey V. and Weides, Martin},
  journal = {Phys. Rev. Lett.},
  volume = {123},
  issue = {19},
  pages = {190502},
  numpages = {6},
  year = {2019},
  month = {Nov},
  publisher = {American Physical Society},
  doi = {10.1103/PhysRevLett.123.190502},
  url = {https://link.aps.org/doi/10.1103/PhysRevLett.123.190502}
}

@article{chen2025etuning,
  title={Scalable and Site-Specific Frequency Tuning of Two-Level System Defects in Superconducting Qubit Arrays},
  author={Chen, Larry and Lee, Kan-Heng and Liu, Chuan-Hong and Marinelli, Brian and Naik, Ravi K and Kang, Ziqi and Goss, Noah and Kim, Hyunseong and Santiago, David I and Siddiqi, Irfan},
  journal={arXiv preprint arXiv:2503.04702},
  year={2025}
}

@article{burnett2019decoherence,
  title={Decoherence benchmarking of superconducting qubits},
  author={Burnett, Jonathan J and Bengtsson, Andreas and Scigliuzzo, Marco and Niepce, David and Kudra, Marina and Delsing, Per and Bylander, Jonas},
  journal={npj Quantum Information},
  volume={5},
  number={1},
  pages={54},
  year={2019},
  publisher={Nature Publishing Group UK London}
}

@article{bilmes2017electronic,
  title={Electronic decoherence of two-level systems in a Josephson junction},
  author={Bilmes, Alexander and Zanker, Sebastian and Heimes, Andreas and Marthaler, Michael and Sch{\"o}n, Gerd and Weiss, Georg and Ustinov, Alexey V and Lisenfeld, J{\"u}rgen},
  journal={Physical review B},
  volume={96},
  number={6},
  pages={064504},
  year={2017},
  publisher={APS}
}

@article{lisenfeld2023enhancing,
  title={Enhancing the coherence of superconducting quantum bits with electric fields},
  author={Lisenfeld, J{\"u}rgen and Bilmes, Alexander and Ustinov, Alexey V},
  journal={npj Quantum Information},
  volume={9},
  number={1},
  pages={8},
  year={2023},
  publisher={Nature Publishing Group UK London}
}

@article{dane2025performance,
  title={Performance Stabilization of High-Coherence Superconducting Qubits},
  author={Dane, Andrew and Balakrishnan, Karthik and Wacaser, Brent and Hung, Li-Wen and Mamin, HJ and Rugar, Daniel and Shelby, Robert M and Murray, Conal and Rodbell, Kenneth and Sleight, Jeffrey},
  journal={arXiv preprint arXiv:2503.12514},
  year={2025}
}

@article{Shalibo2010,
  title = {Lifetime and Coherence of Two-Level Defects in a Josephson Junction},
  author = {Shalibo, Yoni and Rofe, Ya'ara and Shwa, David and Zeides, Felix and Neeley, Matthew and Martinis, John M. and Katz, Nadav},
  journal = {Phys. Rev. Lett.},
  volume = {105},
  issue = {17},
  pages = {177001},
  numpages = {4},
  year = {2010},
  month = {Oct},
  publisher = {American Physical Society},
  doi = {10.1103/PhysRevLett.105.177001},
  url = {https://link.aps.org/doi/10.1103/PhysRevLett.105.177001}
}

@article{hegedus2024,
  title={In-situ scanning gate imaging of individual two-level material defects in live superconducting quantum circuits},
  author={Heged{\"u}s, M and Banerjee, R and Hutcheson, A and Barker, T and Mahashabde, S and Danilov, AV and Kubatkin, SE and Antonov, V and de Graaf, SE},
  journal={arXiv preprint arXiv:2408.16660},
  year={2024}
}

@article{Barends13,
   title={Coherent Josephson Qubit Suitable for Scalable Quantum Integrated Circuits},
   volume={111},
   ISSN={1079-7114},
   url={http://dx.doi.org/10.1103/PhysRevLett.111.080502},
   DOI={10.1103/physrevlett.111.080502},
   number={8},
   journal={Physical Review Letters},
   publisher={American Physical Society (APS)},
   author={Barends, R. and Kelly, J. and Megrant, A. and Sank, D. and Jeffrey, E. and Chen, Y. and Yin, Y. and Chiaro, B. and Mutus, J. and Neill, C. and O’Malley, P. and Roushan, P. and Wenner, J. and White, T. C. and Cleland, A. N. and Martinis, John M.},
   year={2013},
   month=aug }

@article{martinis2005decoherence,
  title={Decoherence in Josephson qubits from dielectric loss},
  author={Martinis, John M and Cooper, Ken B and McDermott, Robert and Steffen, Matthias and Ansmann, Markus and Osborn, KD and Cicak, Katarina and Oh, Seongshik and Pappas, David P and Simmonds, Raymond W and others},
  journal={Physical review letters},
  volume={95},
  number={21},
  pages={210503},
  year={2005},
  publisher={APS}
}

@phdthesis{Sank,
	Type = {PhD thesis},
	Title = {Fast, accurate state measurement in superconducting qubits},
	Author = {Daniel Thomas Sank},
	School = {University of California, Santa Barbara},
	URL = {},
	pages = {229},
	Year = {2014},
}

@article{osman2023mitigation,
  title={Mitigation of frequency collisions in superconducting quantum processors},
  author={Osman, Amr and Fern{\'a}ndez-Pend{\'a}s, Jorge and Warren, Christopher and Kosen, Sandoko and Scigliuzzo, Marco and Frisk Kockum, Anton and Tancredi, Giovanna and Fadavi Roudsari, Anita and Bylander, Jonas},
  journal={Physical Review Research},
  volume={5},
  number={4},
  pages={043001},
  year={2023},
  publisher={APS}
}

@article{neeley2008process,
  title={Process tomography of quantum memory in a Josephson-phase qubit coupled to a two-level state},
  author={Neeley, Matthew and Ansmann, Markus and Bialczak, Radoslaw C and Hofheinz, Max and Katz, Nadav and Lucero, Erik and O’connell, A and Wang, Haohua and Cleland, Andrew N and Martinis, John M},
  journal={Nature Physics},
  volume={4},
  number={7},
  pages={523--526},
  year={2008},
  publisher={Nature Publishing Group UK London}
}

@article{Bilmes_2021,
   title={In-situ bandaged Josephson junctions for superconducting quantum processors},
   volume={34},
   ISSN={1361-6668},
   url={http://dx.doi.org/10.1088/1361-6668/ac2a6d},
   DOI={10.1088/1361-6668/ac2a6d},
   number={12},
   journal={Superconductor Science and Technology},
   publisher={IOP Publishing},
   author={Bilmes, Alexander and Händel, Alexander K and Volosheniuk, Serhii and Ustinov, Alexey V and Lisenfeld, Jürgen},
   year={2021},
   month=oct, pages={125011} }

@Article{Lisenfeld2015,
	Title                    = {Observation of directly interacting coherent two-level systems in an amorphous material},
	Author                   = {Lisenfeld, J\"urgen and Grabovskij, Grigorij J. and M\"uller, Clemens and Cole, Jared H. and Weiss, Georg and Ustinov, Alexey V.},
	Journal                  = {Nature Comm.},
	Pages					= {6182},
	Year                     = {2015},
	Volume                   = {6},
}

@article{meissner2018probing,
  title={Probing individual tunneling fluctuators with coherently controlled tunneling systems},
  author={Mei{\ss}ner, Saskia M and Seiler, Arnold and Lisenfeld, J{\"u}rgen and Ustinov, Alexey V and Weiss, Georg},
  journal={Physical Review B},
  volume={97},
  number={18},
  pages={180505},
  year={2018},
  publisher={APS}
}

@article{Moskalev2022,
	title={Optimization of shadow evaporation and oxidation for reproducible quantum Josephson junction circuits},
	author={Moskalev, Dmitry O and Zikiy, Evgeniy V and Pishchimova, Anastasiya A and Ezenkova, Daria A and Smirnov, Nikita S and Ivanov, Anton I and Korshakov, Nikita D and Rodionov, Ilya A},
	journal={Scientific Reports},
	volume={13},
	number={1},
	pages={4174},
	year={2023},
	publisher={Nature Publishing Group UK London}
}

@article{Hung_2022,
   title={Probing Hundreds of Individual Quantum Defects in Polycrystalline and Amorphous Alumina},
   volume={17},
   ISSN={2331-7019},
   url={http://dx.doi.org/10.1103/PhysRevApplied.17.034025},
   DOI={10.1103/physrevapplied.17.034025},
   number={3},
   journal={Physical Review Applied},
   publisher={American Physical Society (APS)},
   author={Hung, Chih-Chiao and Yu, Liuqi and Foroozani, Neda and Fritz, Stefan and Gerthsen, Dagmar and Osborn, Kevin D.},
   year={2022},
   month=mar }

@article{bilmes2021sensors,
	title={Quantum sensors for microscopic tunneling systems},
	author={Bilmes, Alexander and Volosheniuk, Serhii and Brehm, Jan David and Ustinov, Alexey V and Lisenfeld, J{\"u}rgen},
	journal={npj Quantum Information},
	volume={7},
	number={1},
	pages={27},
	year={2021},
	publisher={Nature Publishing Group UK London}
}

@article{biznarova2024mitigation,
	title={Mitigation of interfacial dielectric loss in aluminum-on-silicon superconducting qubits},
	author={Bizn{\'a}rov{\'a}, Janka and Osman, Amr and Rehnman, Emil and Chayanun, Lert and Kri{\v{z}}an, Christian and Malmberg, Per and Rommel, Marcus and Warren, Christopher and Delsing, Per and Yurgens, August and others},
	journal={npj Quantum Information},
	volume={10},
	number={1},
	pages={78},
	year={2024},
	publisher={Nature Publishing Group UK London}
}

@article{daum2025mergemon,
	title={Investigation of Parasitic Two-Level Systems in Merged-Element Transmon Qubits},
	author={Daum, Etienne and Berlitz, Benedikt and Deck, Steffen and Ustinov, Alexey V and Lisenfeld, J{\"u}rgen},
	journal={arXiv preprint arXiv:2509.22593},
	year={2025}
}

@article{Quintana_2014,
   title={Characterization and reduction of microfabrication-induced decoherence in superconducting quantum circuits},
   volume={105},
   ISSN={1077-3118},
   url={http://dx.doi.org/10.1063/1.4893297},
   DOI={10.1063/1.4893297},
   number={6},
   journal={Applied Physics Letters},
   publisher={AIP Publishing},
   author={Quintana, C. M. and Megrant, A. and Chen, Z. and Dunsworth, A. and Chiaro, B. and Barends, R. and Campbell, B. and Chen, Yu and Hoi, I.-C. and Jeffrey, E. and Kelly, J. and Mutus, J. Y. and O’Malley, P. J. J. and Neill, C. and Roushan, P. and Sank, D. and Vainsencher, A. and Wenner, J. and White, T. C. and Cleland, A. N. and Martinis, John M.},
   year={2014},
   month=aug }

@article{sarabi2016projected,
  title={Projected dipole moments of individual two-level defects extracted using circuit quantum electrodynamics},
  author={Sarabi, Bahman and Ramanayaka, Aruna N and Burin, Alexander L and Wellstood, Frederick C and Osborn, Kevin D},
  journal={Physical review letters},
  volume={116},
  number={16},
  pages={167002},
  year={2016},
  publisher={APS}
}

@article{martinis2021,
	title={Surface loss calculations and design of a superconducting transmon qubit with tapered wiring},
	author={Martinis, John M},
	journal={npj Quantum Information},
	volume={8},
	number={1},
	pages={26},
	year={2022},
	publisher={Nature Publishing Group UK London}
}
\onecolumngrid

\clearpage
\noindent \LARGE{Mapping the positions of Two-Level-Systems on the surface of a superconducting transmon qubit}\\[0.2cm]
\normalsize{J\"urgen Lisenfeld, Alexander K. H\"andel, Etienne Daum, Benedikt Berlitz, Alexander Bilmes, and Alexey V. Ustinov}\\

\section*{Supplementary material}
\renewcommand{\thefigure}{Supplementary Figure~\arabic{figure}}
 \renewcommand{\thepage}{S\arabic{page}}
 \renewcommand{\thesection}{Supplementary Note \Alph{section}}
 \renewcommand{\thetable}{S\arabic{table}}
 \renewcommand{\thefigure}{S\arabic{figure}}
 \renewcommand{\figurename}{Fig.}
 \setcounter{figure}{0}
 \setcounter{section}{0}

\renewcommand{\theequation}{A\arabic{equation}}
\setcounter{equation}{0} 

\section{\label{app:sample}Sample fabrication and properties}
\noindent The qubit sample investigated in this work was fabricated with the following recipe.\\

\noindent \textbf{Wafer preparation.} Sapphire wafers of 3 inch diameter and 500 \textmu m thickness are cleaned during 10 minutes in piranha solution and an additional exposure to $O_2$-plasma for 10 minutes before loading them into a Plassys eBeam-evaporator. During the following night, the Plassys is pumped and the wafer is tempered at 200\degreecelsius\ for 2 hours.\\
The wafer is then coated by 100\,nm of aluminum at a typical background pressure of 1.4$\cdot 10^{-7}$ mBar. Before the coated wafer is exposed to atmosphere, it is oxidized for 10 minutes at 30 mBar $O_2$ pressure.\\

\textbf{Dicing.} The 3-inch wafers are coated with a S1818 protective resist layer of $\approx$1.8 \textmu m thickness and diced into 2cm x 2cm large pieces. The diced wafers are cleaned in DMSO at 90\degreecelsius \ without ultrasonic excitation to avoid damage of the aluminum layer by sapphire fragments from the dicing step.\\

\textbf{Optical lithography.} The cleaned 2x2 cm wafers are coated with a 170 \textmu m-thick layer of S1805 photoresist. The larger structures such as bonding pads, ground plane, resonators, qubit island, flux lines, and gate electrodes, are patterned via a photomask at a light intensity of 1.9 mW/cm$^2$ during 10 seconds and development in a 3:2 AZ-Developer and water mixture for about 30s. The aluminum layer is then dry-etched in an ArClO plasma (15 sccm Ar, 3sccm Cl, 1sscm O$_2$) for 1 minute, and developed in DMSO at 90\degreecelsius\  for 2-3 hours immediately after etching in order to prevent chlorine-induced corrosion.\\

\textbf{EBeam lithography.} The junctions are patterned by 3-angle shadow evaporation based on a Dolan-bridge process where unwanted stray junctions are shorted as explained in Ref.~\onlinecite{Bilmes_2021}. To reduce aging effects, the wafer is afterwards oxidized in the evaporator's vacuum chamber in a clean Oxygen atmosphere of 30 mBar pressure. Final cleaning of the wafers occurs in DMSO at 90\degreecelsius\ with ultrasonic excitation.\\

\textbf{Chip dicing.} The wafers are finally diced into chips of 6x6 mm size after additional coating with protective resist as described above.\\

Figure S1 shows a photograph of a finished chip. 
A summary of the qubit circuit parameters is given by Table~S1.\\

\vspace{0.5cm}
\begin{minipage}[h!]{0.3\textwidth}
		\includegraphics[width=1.3\linewidth]{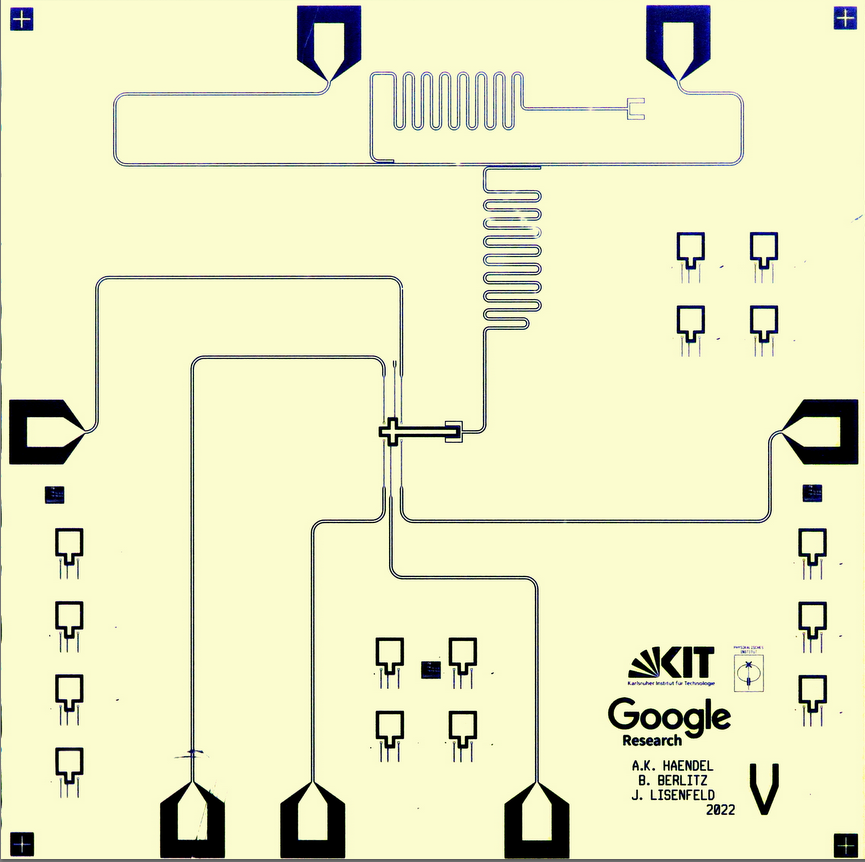}
						
		Fig. S1. Photograph of the qubit chip for TLS mapping.
\end{minipage} 
\hspace{2cm}
\begin{minipage}[h!]{0.55\textwidth}
	\begin{tabular}{llll}
		\hline
		name & value & description\\ 
		\hline 
		$f_\mathrm{res,max}$ & 7.83 GHz & resonator frequency at max. qubit frequency\\ 
		$f_\mathrm{q,max}$ & 5.56 GHz & maximum qubit frequency at zero flux\\ 
		$T_{1,max}$ & 8 \textmu s& observed maximum qubit $T_1$ time \\ 
		$T_{1}$ & 5 \textmu s& qubit $T_1$-time averaged over frequency\\  
		\hline
		$C_{q}$ & 84 fF & capacitance of qubit island to ground \\ 
		$C_{q,\alpha}$ & 0.24 fF & capacitance of qubit island to $\alpha$-electrode \\ 
		$C_{q,\beta}$ & 0.29 fF & capacitance of qubit island to $\beta$-electrode \\ 
		$C_{q,\gamma}$ & 0.24 fF & capacitance of qubit island to $\gamma$-electrode \\ 
		$C_{q,\delta}$ & 0.29 fF & capacitance of qubit island to $\delta$-electrode \\ 
		\hline
		$R_{n,\mathrm{JJ}}$ & 14.5 k$\Omega$ & room-temperature resistance of a single JJ\\
		$R_{n,\mathrm{SQ}}$ & 7.25 k$\Omega$ & room-temperature resistance of a 2-JJ SQUID\\ 
		$I_{c,\mathrm{JJ}}$  & 23 nA & single-JJ critical current\\     
		$A_\mathrm{JJ}$ & 0.08 $\micro \mathrm{m}^2$ & JJ area (for size 260$\pm 10$ nm $\cdot$ 310$\pm 10$ nm)\\
		$j_{c}$  & 285 nA/$\micro \mathrm{m}^2$ & critical current density\\ 
		\hline    
		$E_{c}/h$  & 229 MHz & qubit charging energy\\ 
		$E_{J}/h$  & 23.4 GHz & qubit Josephson energy\\     
		\hline
	\end{tabular}
	
	 \vspace{0.3cm} Table S1. Measured and simulated fabrication parameters of the studied qubit sample. Capacitance values were extracted with Ansys Maxwell. Junction (JJ) and SQUID resistances are averages from 4-point measurements at room temperature. 
	\label{table:fitparms}
	\label{fig:app-chip}
\end{minipage}\\

\vspace{0.5cm}

\section{\label{app:setup}Experimental setup}
\noindent
\begin{minipage}[b]{0.3\textwidth}
\vspace{0pt}
Figure S2 illustrates the experimental setup with the components for dispersive qubit readout, qubit flux biasing, and supply of voltage bias to the four on-chip gate electrodes.\\
The voltages for the on-chip gate electrodes are generated by a 16-bit DAC that is controlled via an optical connection to a PC, and then amplified to a range of +/- 250 V using piezo drivers. The signals are transmitted via a copper wire loom to the qubit chip after passing through an LCR low-pass filter at the 4K-temperature stage. All measurements were done at a sample temperature of 25 - 30 mK.\\
\end{minipage}
\begin{minipage}{0.7\textwidth}
	\vspace{-3.5cm}
\hspace{1.8cm}
	\includegraphics[width=0.8\linewidth]{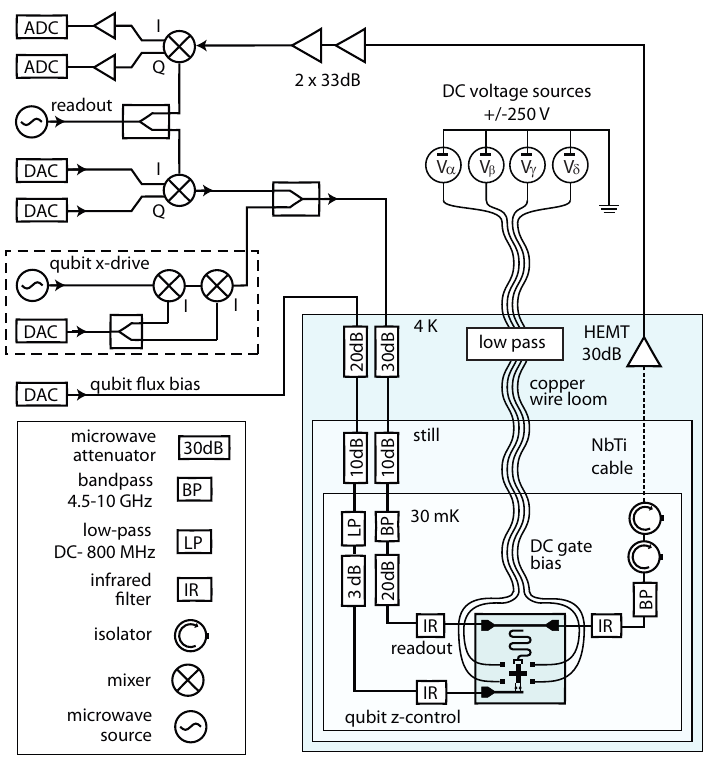}\\
\phantom{hi there}\small Fig. S2. Schematic of the experimental setup. 
\end{minipage}
\phantom{hi there}\\

\twocolumngrid
\section{Loss from gate electrodes}\label{app:loss}
\noindent The qubit's energy relaxation rate is enhanced by its capacitive coupling to the gate electrodes. In order to increase the spatial resolution in TLS mapping, the electrodes have been placed in close vicinity to the qubit island as can be seen in the design shown in Figs.~\ref{fig:1}b and c. The resulting coupling capacitance between the qubit island and each gate electrode is $C_\mathrm{q,i}\approx 0.24 - 0.29$~fF as simulated with Ansys Maxwell, while the capacitance of the qubit island to ground is $C_q =$84~fF. This results in a loaded quality factor of the qubit given by~\cite{Sank,lisenfeld2023enhancing}
\begin{equation}
Q_l = \left ( \frac{C_\mathrm{tot}}{C_c}\right )^2 \frac{Z_q}{\mathrm{Re}(Z_\mathrm{eff})},
\end{equation}\\
where $C_\mathrm{tot} = C_q + \sum_i C_{q,i}$ is the total capacitance, and $Z_q = \sqrt{L_q / C_\mathrm{tot}} \approx 290\,\Omega$ is the qubit impedance estimated from the inductance of its two parallel Josephson junctions $L_q = 0.5\cdot L_J = 0.5\cdot h / 4\pi e I_c$ each having a critical current of $I_c\approx 23$\,nA as shown in Table~\ref{table:fitparms}. Assuming a real part impedance of $\mathrm{Re}(Z_\mathrm{eff})\approx 50\,\Omega$ for the coplanar gate electrode including its feed line results in $Q_l\approx 0.6$ million. The resulting loaded quality factor for all four electrodes is $Q_\mathrm{tot}=(4/Q_l)^{-1}$ and limits the qubit's energy relaxation time to $T_1 = Q_\mathrm{tot}/2\pi f_q \approx 5\,$\textmu s at a qubit resonance frequency of $f_q = 5$\,GHz.\\
This estimated upper limit of $T_1$ is in the range of the observed average qubit relaxation time $T_1\approx 8\,$\textmu s. For comparison, equally fabricated qubit samples with comparable design but without gate electrodes showed $T_1$ times between 10 and 20 \textmu s. The actual impact of the gate electrodes on the $T_1$ time remains unclear due to uncertainty about the  gate electrode impedance, and the fact that only a single sample with on-chip gate electrodes was measured. To reduce this loss, the gate electrode impedance can be adjusted using on-chip shunt capacitors or inductors~\cite{bilmes2021sensors}.\\

\section{\label{app:simus} Electric field simulations}

\new{
\noindent For the TLS mapping procedure and the energy-participation-ratio (EPR) analysis, the AC-electric field distribution of the transmon qubit was simulated using \textit{ANSYS HFSS}'s Eigenmode Solver. The DC-electric fields emitted by the four gate electrodes were simulated with \textit{ANSYS MAXWELL}, separately biasing each electrode with 1V. In all models, the simulation volume was restricted to a 1mm$\times$1mm$\times$1mm large region centered around the qubit. All thin-film structures, including the qubit island, Josephson junctions, gate electrodes, resonator, bias line, and ground plane, were modeled as 2D sheets on a 1mm$\times$1mm$\times$0.5mm sapphire substrate. Except for the junctions, all sheets were assigned the ``Perfect E'' boundary condition, while the junction sheets were assigned the ``Lump RLC'' boundary condition. Assuming a relative permittivity $\epsilon_{\mathrm{r}} = 10$ and thickness d = 2nm of the junction barrier, the applied junction capacitances $C_{\mathrm{JJ}}$ were estimated from the known junction area A, using $C_{\mathrm{JJ}} = A\epsilon_{0}\epsilon_{\mathrm{r}}/d$, to approximately 3.6~fF. The junction inductances were adjusted such that the simulated eigenfrequency of the transmon qubit was approximately $f_{\mathrm{q}} \approx$ 5~GHz. Along the edges of qubit sheets and the entire SQUID loop, the precision of the adaptive mesh solver was set to a minimum of 100~nm. Finally, the energy of the qubit mode was normalized to one photon to obtain the electric field strength $\erms$.\\  }
 
 \new{   
 Systematic offsets in the simulated electric field strength, e.g. due to averaging over simulation mesh elements, play a minor role since only relative field strengths are considered in the method. We used the same mesh for all simulations and increased the meshing resolution in a 2~\textmu m wide region around electrode edges. Systematic errors however do affect the estimation of the TLS dipole moments, which are directly calculated from the simulated local DC-field at the deduced TLS position and the measured TLS tuning strength. The steep gradient of $\erms$ near electrode edges enhances the uncertainty of the TLS dipole moment. Despite this, the obtained TLS dipole moments (Fig. 4b in the main manuscript) have a reasonable mean value and do not show excessive spread compared to other results~\cite{Hung_2022}.\\  
  }

\section{\label{app:minfield} TLS detectability limits}
\noindent
As described in the main text, TLS can only be detected in regions where the AC electric field of the qubit mode is above a minimum strength $E_\mathrm{min}$. In this case, the TLS-qubit coupling energy exceeds $g_\mathrm{min} = p\,E_\mathrm{min}$, and when the qubit is in resonance with the TLS, the qubit $T_1$ time is sufficiently suppressed so that the TLS is observed. Here, $p$ is the component of the TLS' electric dipole moment that is parallel to the qubit's AC electric field, whose value can be estimated to $p_\parallel \approx 1 e$\AA \ in accordance to measurements\cite{Martinis2005,Barends13,Shalibo2010,Lisenfeld19,Lisenfeld2015,chen2024,Hung_2022} and atomistic simulations\cite{paz2014,dubois20153d}.\\
When a moderately coupled TLS is near resonance with the qubit, it causes a Lorentzian peak in the qubit's energy relaxation rate $\Gamma_{1} \equiv 1/T_1$, which in the limit of $\Gamma_\mathrm{1,TLS} > g/h > \Gamma_{1,q}$ is described (see supplementary material to \cite{klimov2018fluctuations})
\begin{equation}
    \Gamma_{1} = \frac{2\,(g/h)^2\,\Gamma}{(\Gamma/2\pi)^2 + \delta^2}+\Gamma_\mathrm{1,q}.
    \label{eq:decrate}
\end{equation}
Here, $\Gamma = \Gamma_{1,\mathrm{TLS}}/2 + \Gamma_{2,\mathrm{TLS}} + \Gamma_{1,\mathrm{q}}/2 + \Gamma_{2,\mathrm{q}}$ is the sum of TLS and qubit energy relaxation and dephasing rates, and $\delta$ is the detuning between qubit and TLS.\\

Solving for the coupling strength $g$ where Eq.~(\ref{eq:decrate}) predicts that the qubit's energy relaxation time at resonance $\delta=0$ is decreased by a factor $\kappa$ then results in the minimum coupling strength $g_\mathrm{min} = \hbar \sqrt{\kappa \cdot \Gamma_\mathrm{1,q} \,\Gamma /2}$ and corresponding minimum AC-electric qubit field strength $E_\mathrm{min} = g_\mathrm{min} / p_\parallel$.\\
Figure~S3 shows $E_\mathrm{min}$ vs. the detection factor $\kappa$ for different TLS dipole moment sizes $p_\parallel$ and a qubit $T_1$ time of 7 \textmu s as in our experiment.\\

\includegraphics[width=1.\linewidth]{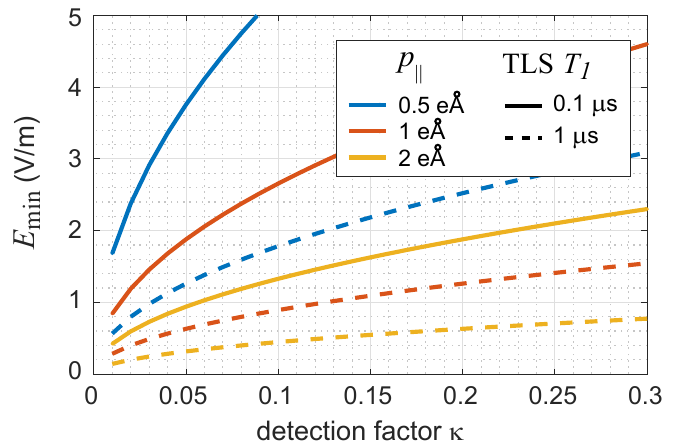}
\small{Fig. S3. Minimum qubit AC-electric field strength $E_\mathrm{min}$ at the TLS position for which the qubit $T_1$-time is reduced by more than the detection factor $\kappa$ at resonance. For this plot, the isolated qubit $T_1$ time is set to 7 $\mu$s. The colors correspond to different TLS electric dipole moments. Solid and dashed lines are plotted for TLS $T_1$ times of 0.1 \textmu s and 1 \textmu s, respectively.}\\

The minimum field $E_\mathrm{min}$ decreases for more coherent TLS. In experiments, TLS energy relaxation times of $T_\mathrm{1,TLS} \approx 0.1 - 0.2$ \textmu s are most commonly observed~\cite{Barends13,Shalibo2010,Lisenfeld19,Lisenfeld2015,Bilmes_2021,chen2024}. However, also TLS are observed which have longer coherence times in a range of a few 10 \textmu s. These may be better decoupled from the phonon bath~\cite{chen2024}, for example due to a symmetry of the tunneling wavefunction, or because of a reduced local phonon spectrum e.g. at material interfaces.\\

The region of a qubit circuit within which TLS can be detected is bounded by $E_\mathrm{min}$ and increases with the qubit coherence time $T_1$ as shown in Fig. S4. More coherent qubits are thus affected by a larger number of TLS, including those residing farther away from the electrode edges.\\

\includegraphics[width=0.8\linewidth]{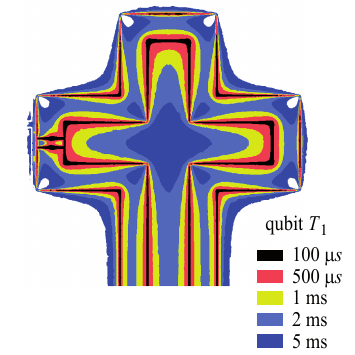}\\
\small Fig. S4. The region where TLS interact sufficiently strongly to be detected in TLS spectroscopy depends on the $T_1$ time of the qubit, as indicated by the different colors. The plot was made with an Ansys HFFS simulation of the qubit's AC electric field strength, assuming a TLS coherence time of 100ns, a TLS dipole moment of 1 e\AA , and a detection factor of $\kappa=0.1$. \normalsize \\	


\new{\textbf{Energy participation ratio (EPR).} The probability to detect a TLS increases with the square of the local qubit AC-field, since the qubit decay rate $\Gamma_1$, which is used as the signal in swap spectroscopy, scales with the qubit-TLS coupling strength $g^2\propto |\erms|^2$ (see Eq.~\ref{eq:decrate} and definition of $g$ in the main text). This is analogous to surface loss analyses which consider the energy participation ratio (EPR) of different circuit areas, i.e. the fraction of the qubit energy that is stored there~\cite{wang2025superconducting}. We thus estimate the expected ratio of the numbers of TLS observed on the SQUID vs. those on the qubit capacitor by comparing integrals over $|\erms|^2$ in the vicinity of these elements.\\
	\indent For this, the simulated $|\erms|$ is exported on a rectangular grid, and the sum of $|\erms|^2$ is calculated for pixels inside the SQUID and capacitor areas whose dimensions are shown in Fig.~\ref{fig_app_mask}. Moreover, this sum is limited to regions where $|\erms|>E_\mathrm{min}$, in order to account for the detectability of TLS in swap spectroscopy. The resulting EPRs of SQUID and ground+island are shown in the main text's Fig.~4c as a function of the chosen value of $E_\mathrm{min}$.\\
}
 \setcounter{figure}{4}

	\begin{figure}
	\includegraphics[width=0.6\linewidth]{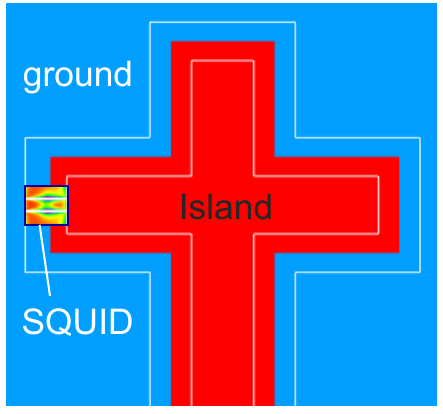}\\
\caption{Areas assigned to the ground plane (blue), the qubit island (red), and the SQUID (indicated square), as used for calculating energy participation ratios. The white lines indicate electrode edges.}
\label{fig_app_mask}
	\end{figure}

\new{\textbf{Field directions.} The method to find TLS positions requires that the gate electrode fields are parallel at the position of the TLS. This is confirmed with Fig.~\ref{fig:app-vectors} also for the region near the Josephson junctions. Fig.~\ref{fig:app-vectors}a shows the strength of the qubit's AC-electric field $|\erms|$ and a zoom onto the DC-SQUID. A vertical cross-section is shown in Fig.~\ref{fig:app-vectors}b, where the junction leads are indicated by gray bars that are separated by 6~\textmu m. The region where $|\erms| > E_\mathrm{min}$ (here for $E_\mathrm{min}$=0.75 V/m) is shaded red. Directly at the junctions, the detection region extends up to 1.8~\textmu m distance from the lead edges. Similar to Fig.~3 in the main text, the $\vec{x}-$ and $\vec{z}-$directions of the electric fields are plotted in Fig.~\ref{fig:app-vectors}c, and their vector products shown in Fig.~\ref{fig:app-vectors}d indicate that they are practically parallel within the vicinity of junction leads.}\\

 \setcounter{figure}{6}
\begin{figure*}[htb]
	\centering
	\includegraphics[width=1\textwidth]{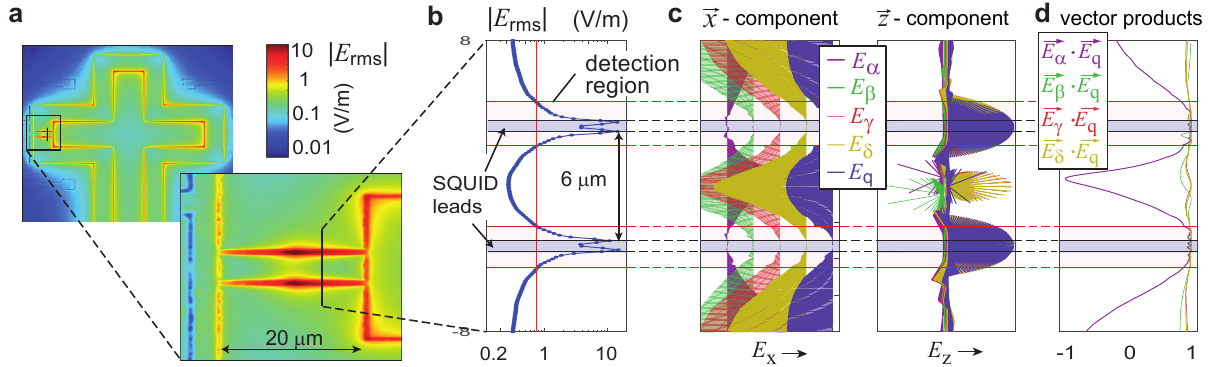}
	\caption{\textbf{Strength and orientations of AC- and DC-electric fields near the DC-SQUID.}
		\textbf{a} Magnitude of the qubit's AC-electric field $|\erms|$ as simulated with Ansys HFSS, with a zoom onto the DC-SQUID. \textbf{b} Cross-section of $|\erms|$ along the black line in \textbf{a}. TLS can be detected in the red-shaded area where the field exceeds a minimum strength $E_\mathrm{min}$. \textbf{c} Components of the electric fields from the four electrodes and the qubit, plotted in $\vec{x}-$ and $\vec{z}$ directions (left and right panel). \textbf{d} The normalized vector product between the fields of gate electrodes and the qubit indicates that they are practically parallel in the range where TLS can be detected.}
	\label{fig:app-vectors}
\end{figure*}

\section{Data acquisition and TLS properties}\label{app:data}
\noindent In the TLS swap-spectroscopy measurements as shown in Fig.~2a, the qubit was swept in a frequency range from 5.05 GHz to 5.20 GHz in steps of 0.375 MHz. In each segment, the voltage of a different gate electrode was increased by 1V in steps of 50 mV. The total voltage on each electrode was swept from -60 V to +100 V to acquire a total number of 640 segments.\\
	
	We have identified the resonant traces of 55 individual TLS whose response to all four gate electrodes was observed in the investigated frequency range, allowing those to be further analyzed. Each segmented hyperbolic trace was manually marked and fit to Eq.~\ref{eqn:myTLShyp} to obtain the TLS' response factors $\gamma_i$, which are plotted in Fig.~S5. We acquired data during 33 days, comprising over 5 million individual (averaged) measurements at a rate of ~2 measurements / second. The measurement duration could be reduced to about 1 day by using faster electronics, a Josephson parametric amplifier, and active qubit reset techniques.\\
	
	\includegraphics[width=0.95\linewidth]{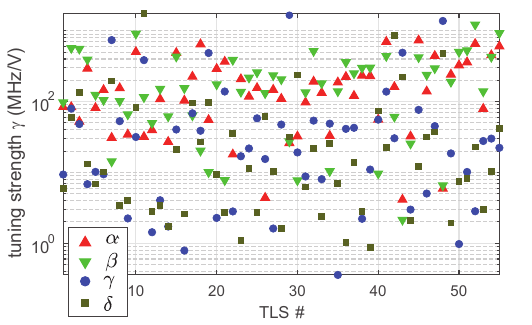}
	\small{Fig. S6. Tuning strengths $\gamma_i$ for the four gate electrodes $i \in \{\alpha,\beta,\gamma,\delta\}$, obtained for 55 TLS by fitting their swap-spectroscopy traces to Eq.~(\ref{eqn:myTLShyp}).
	}

\section{Resolution in TLS mapping}
\label{app:resolution}
\noindent The precision in the determination of TLS positions depends mostly on the uncertainties in the tuning strenghts $\gamma_i$ measured in segmented swap spectroscopy as shown in Fig.~\ref{fig:2}a. If a TLS is located at a larger distance from an electrode, its weaker response results in a less precise fit estimate. This can be mitigated by a more appropriate qubit and gate electrode design that aims to enhance the overlap of the electric fields.\\

Moreover, TLS which respond strongly to one electrode may be quickly tuned through most of the observed swap spectroscopy range, reducing the available data on the response to more distant electrodes. This problem can be avoided with a measurement protocol where the gate voltages are reset to an initial value after they were swept. This would also allow one to increase the swept voltage range to facilitate measurements of weak responses to some electrodes.\\
As an example of the resolution limit, Fig.~\ref{fig:app-SQUIDTLS} shows data of a TLS that is determined to reside near the qubit's Josephson junctions. There is some uncertainty whether the TLS resides on the upper or lower branch of the DC-SQUID. While most observed tuning ratios indicate a position on the upper branch, the data for $\gamma_\alpha / \gamma_\gamma$ (lower left panel in Fig.~\ref{fig:app-SQUIDTLS}) suggests a solution on the lower branch. This may be due to the fit error determining the relatively weak $\gamma$-gate response $\gamma_\gamma$. The SQUID branches are separated by a distance of 6~\textmu m (see also Fig.~\ref{fig:1}d). We expect that in our experiment, the achieved resolution is of similar size, and believe that it could be reduced to a few micrometers by mentioned improvements in sample design and measurement protocol.\\

\new{We tested the impact of errors on the measured TLS-electrode coupling strengths $\gamma_i$ with Monte-Carlo-simulations. For this, a TLS position was manually selected, and the exact tuning ratios at this position were determined from the E-field simulation. To simulate measurement uncertainty, these exact ratios were multiplied by random factors that have a constant distribution in an interval of $1 \pm \delta$, where the maximum variation $\delta$ ranges from 1\% to 50\%. From these noisy tuning ratios, the TLS position was then determined with the mapping algorithm.}\\

\new{For a TLS placed on the SQUID and for a TLS on the qubit island, Fig.~\ref{fig:app-rando-SQUID} and Fig.\ref{fig:app-rando-island} respectively show resulting TLS positions from 1000 iterations (panels \textbf{a} and \textbf{d}). The deviations from the exact TLS location are shown in panels \textbf{b} for different variation factors $\delta$, and in panel \textbf{e} for differently chosen cut-off of the minimum electric field $E_\mathrm{min}$ that limits the detection area. We find that the median deviation between estimated and exact TLS positions ranges between 2~\textmu m and 8~\textmu m for variation factors up to $\delta=50$\%, while for factors of $\delta>20\%$, the TLS position on the SQUID can not anymore reliably distinguished from a position on the island or ground plane. When the detection area is not restricted at small values of $E_\mathrm{min}$, TLS are mostly placed further away from the electrodes (see Fig.~\ref{fig:app-rando-SQUID}d). Such solutions are invalid since at these positions, TLS would not be observed in our TLS spectroscopy, and because the electric fields from the four electrode are not necessarily parallel, e.g. in the gap between ground plane and qubit island.}\\

\setcounter{figure}{7}
\begin{figure*}[hb]
	\centering
	\includegraphics[width=0.85\textwidth]{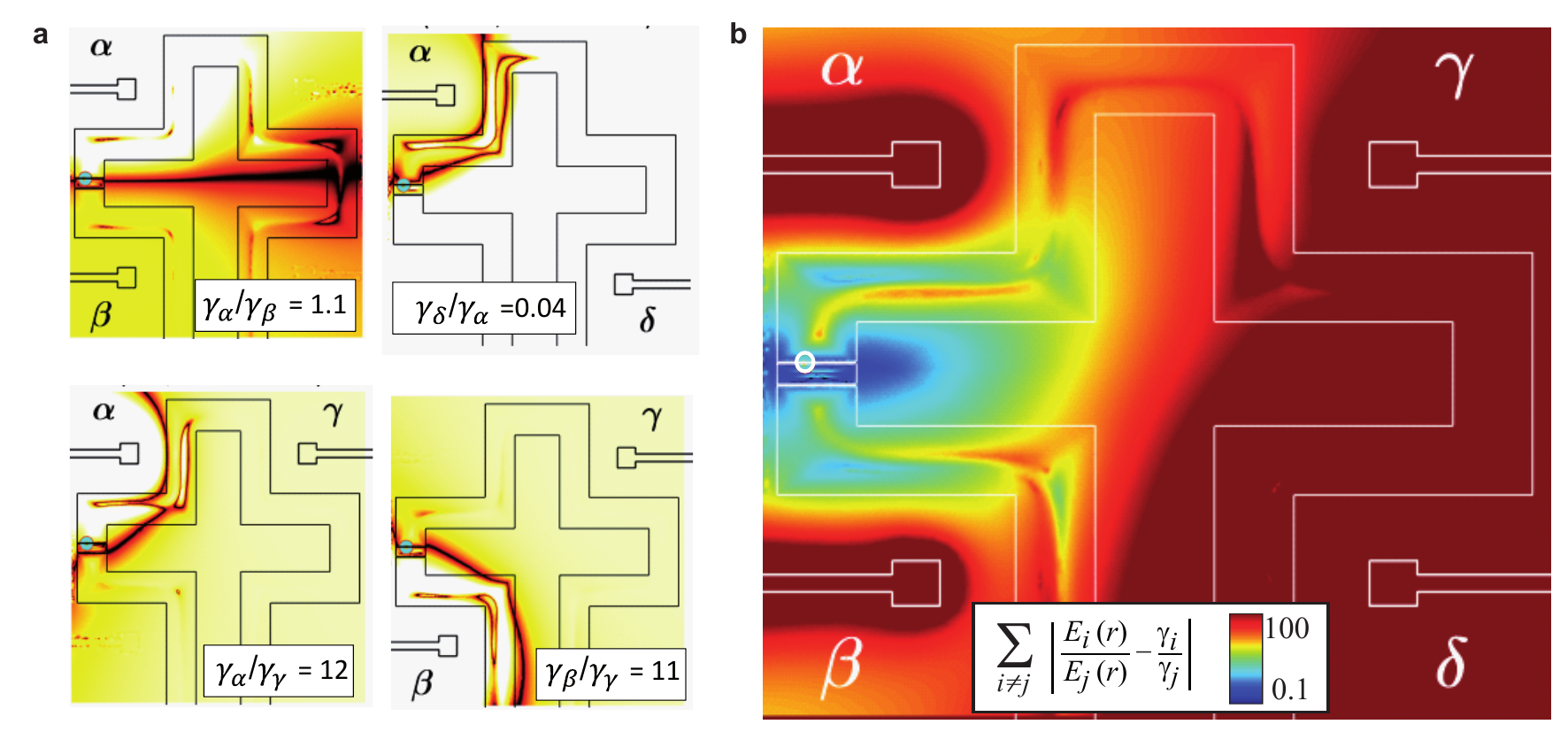}
		\caption{Data of a TLS that is identified to reside on the qubit's DC-SQUID, similar to Figs.~\ref{fig:2}b,c. \textbf{a}   
		\textbf{b}
		Difference sum $\sigma$ (Eq.~3) over all 6 unique combinations of electrode pairs. The white circle marks the global minimum, placing the TLS on the upper branch of the DC-SQUID's loop. 	}		
	\label{fig:app-SQUIDTLS}
\end{figure*}

\begin{figure*}[hb]
	\centering
	\includegraphics[width=\textwidth]{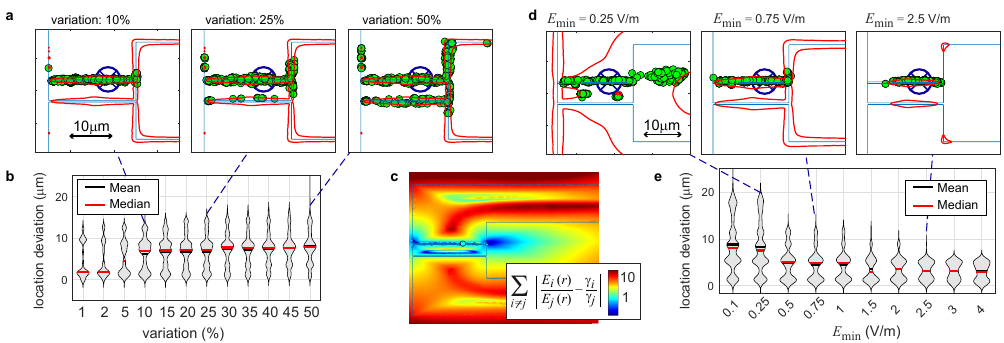}
\caption{\textbf{a} Determined positions (green circles) of a TLS on the DC-SQUID, whose exact location is marked by the blue circle, for different levels of noise (variation factor $\delta$) in the determined tuning ratios. \textbf{b} Histograms of deviations from the exact TLS position for different levels of variation $\delta$. 
\textbf{c} Difference between simulated E-field strengths and TLS tunability as in Fig.~2c, where minima (blue pixels) indicate most likely TLS positions. 
\textbf{d} Same as \textbf{a}, but varying the value of $E_\mathrm{min}$ that limits the detection area. \textbf{e} Deviations from the exact TLS location vs. $E_\mathrm{min}$. For Figs.~\textbf{a, b}, $E_\mathrm{min}=1$ V/m, and for Figs.~\textbf{d, e}, the variation was fixed at $\delta=10\%$
}
	\label{fig:app-rando-SQUID}
\end{figure*}

\begin{figure*}[hb]
	\centering
	\includegraphics[width=\textwidth]{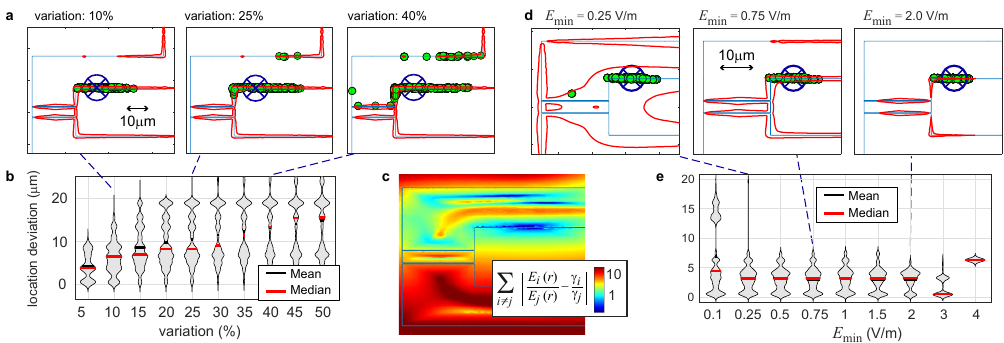}
	\caption{Determined positions (green circles) of a TLS on the qubit island (blue circle), similar to Fig.~\ref{fig:app-rando-SQUID}.
	\textbf{a} and \textbf{d}  Positions for different variation factors $\delta$ and values of $E_\mathrm{min}$, respectively, 		
    \textbf{b} and \textbf{e}: Histograms of deviations from the exact TLS position for different levels of variation $\delta$, and for different values of $E_\mathrm{min}$, respectively. \textbf{c}  Difference between simulated E-field strengths and TLS tunability.}
	\label{fig:app-rando-island}
\end{figure*}

\new{Before analyzing the data, we excluded several detected traces of TLS whose coupling strengths to the upper electrodes ($\alpha$ and $\gamma$ in Fig.~\ref{fig:1}) were indistinguishable from zero so that their position can not be reliably determined with the current procedure. These traces stem from TLS that are located further away on the extended arm of the qubit island (see Fig.~\ref{fig:1}b). Accordingly, possible solutions for TLS positions were restricted to the 105~\textmu m$\times$105~\textmu m-large window as shown in Fig.~\ref{fig:1}c.}\\.

\end{document}